\documentclass[final]{svjour2}
\usepackage{graphicx}
\usepackage{rotating}
\usepackage{amssymb}
\usepackage{mathptmx}
\usepackage[numbers]{natbib}
\makeatletter
\journalname{Journal of Low Temperature Physics}

\bibpunct{}{}{,}{s}{}{,}
\newcommand{\beq}{\begin{equation}}
\newcommand{\eeq}{\end{equation}}
\newcommand{\beqa}{\begin{eqnarray}}
\newcommand{\eeqa}{\end{eqnarray}}
\newcommand{\ba}{\begin{array}}
\newcommand{\ea}{\end{array}}

\begin{document}

\title{Shell Effects in the First Sound Velocity 
of an Ultracold Fermi Gas} 
\author{Luca Salasnich$^{1}$ and Flavio Toigo$^{2}$} 
\institute{$^{1}$CNR-INFM and CNISM, Unit\`a di Padova, 
Via Marzolo 8, 35131 Padova, Italy \\
$^{2}$Dipartimento di Fisica ``Galileo Galilei'' and CNISM, Universit\`a 
di Padova, \\ 
Via Marzolo 8, 35131 Padova, Italy} 

\date{\today}

\maketitle

\keywords{Degenerate Fermi gases, first sound, low-dimensional systems}

\begin{abstract} 
We investigate the first sound of a normal dilute and ultracold 
two-component Fermi gas in a harmonic microtube, i.e. 
a cylinder with harmonic transverse radial confinement in the 
length-scale of microns. 
We show that the velocity of the sound that propagates along the 
axial direction strongly depends on the dimensionality 
of the system. In particular, we predict that 
the first-sound velocity display shell effects:  
by increasing the density, that is 
by inducing the crossover from one-dimension to three-dimensions, 
the first-sound velocity shows jumps in correspondence with 
the filling of harmonic modes. The experimental achievability 
of these effects is discussed by considering $^{40}$K atoms.

PACS numbers: 03.75.Ss, 67.40.Mj, 68.65.-k 
\end{abstract} 


\section{Introduction}

Macroscopic effects of quantum statistics are observable 
at ultra-low temperatures with bosonic vapors of alkali-metal atoms
\cite{book-pethick,book-stringari}. 
In the last years quantum degeneracy 
has been achieved also with fermionic atoms
\cite{demarco,truscott,modugno,greiner,joachim}.  
It is important to observe that 
the role of dimensionality in these 
degenerate gases has been experimentally 
studied until now only with bosons\cite{gorlitz,schreck,kinoshita}.   
For fermions, it has been predicted that a reduced dimensionality  
strongly modifies density profiles\cite{schneider,sala0,sala1,vignolo},   
collective modes\cite{minguzzi} and stability of mixtures
\cite{das,sala2}. Sound velocity has been theoretically 
investigated in strictly one-dimensional (1D) and 2D 
configurations with both normal\cite{minguzzi,capuzzi1} 
and superfluid Fermi gases\cite{ghosh,capuzzi2,capuzzi3}. 
\par 
In this paper we analyze the first-sound 
velocity of a normal Fermi gas in the 1D-3D crossover. 
We stress that in previous studies
\cite{sala-reduce1,sala-reduce2} the sound velocity 
of a Bose gas was considered in the same dimensional crossover. 
Here we consider a dilute and ultracold Fermi gas 
with a strong harmonic confinement along two directions
and uniform in the other direction. 
We show that the first-sound velocity 
strongly depends on the dimensionality of the system. 
We reproduce the results of Minguzzi {\it et al.}\cite{minguzzi} 
and of Capuzzi {\it et al.}\cite{capuzzi1} 
as limiting values. In addition, we predict that 
the sound velocity display observable shell effects. 
By increasing the density, that is
by inducing the crossover from one-dimension to three-dimensions,
the sound velocity show jumps in correspondence with
the filling of harmonic modes shells. 

\section{Confined Fermi gas in a harmonic microtube: 1D-3D crossover}  

Collective sound modes in  a normal Fermi system may be 
classified into two regimes according to whether collisions 
between atoms are important or not.
If the atomic collision time $\tau_c$ 
is much longer than the period ($\omega^{-1}$) of the mode,
i.e. if $\omega \, \tau_c \gg 1$, collisions may be neglected 
and the mode, called zero sound, describes the propagation 
of a deformation of the Fermi sphere \cite{landau,pines,fetter,negele}.  
If, on the contrary, collisions between atoms are so frequent, 
and therefore $\tau_c$ so small, that  $\omega \, \tau_c \ll 1$, 
then the system is in the collisional (or hydrodynamic) 
regime, and the mode is the ordinary (first) sound 
corresponding to a the propagation of a density 
fluctuation \cite{landau,pines,fetter,negele}. 
Here we analyze the collisional case of a normal 
Fermi gas in a harmonic microtube, i.e. a cylinder 
with a harmonic radial confinement with a characteristic length 
of the order of some microns. 
We investigate the regime where the temperature $T$ of the gas 
is well below its Fermi temperature $T_F$, i.e. we consider
the hydrodynamic regime of an otherwise perfect 
and completely degenerate Fermi gas \cite{bruun}.  

Let us consider the Fermi gas   
confined by a harmonic potential of frequency $\omega_{\bot}$ 
in the $x-y$ plane, while it is free to move along the $z$ axis. 
The external potential is then:
\beq 
U({\bf r}) = {1\over 2} m \omega_{\bot}^2\, (x^2+y^2)  \; ,
\eeq
By imposing periodic boundary conditions in a box of length $L$ along $z$, 
the single-particle energies are:
\beq
\epsilon_{i_x i_y i_z} =
\hbar \omega (i_x + i_y + 1) + {\hbar^2 \over 2m}
{(2\pi)^2 \over L^2} i_z^2
\; , 
\label{energy1}
\eeq
where $i_x$, $i_y$ are natural numbers
and $i_z$ is an integer. 
The total number of particles is given by:
\beq 
N = 2\, \sum_{i_x i_y i_z} 
\Theta( \bar{\mu} - \epsilon_{i_x i_y i_z} ) \; ,   
\eeq
where the factor $2$ takes account of spins, 
$\Theta(x)$ is the Heaviside step function, and 
$\bar{\mu}$ is the chemical potential (Fermi energy).  
In a cigar geometry, with $L\gg 2\pi\hbar/\sqrt{2m \bar{\mu}}$, 
one may define the quasi-continuum variable $k_z = {2\pi\over L} i_z$ and
rewrite the number of particles  as 
\beq
N = 2\, \sum_{i_x i_y} {L\over 2\pi} 
\int dk_z \, \Theta( \bar{\mu} - \epsilon_{i_x i_y k_z} ) \; . 
\eeq
Since the motion is free along this direction, the 1D density $n_1={N\over L}$ 
does not depend on $z$,
 and it can be written as 
\beq 
n_1 = {1\over \pi} 
\sum_{i_x i_y} \int dk_z  \, 
\Theta( \bar{\mu} - \epsilon_{i_x i_y k_z} ) 
= {1\over \pi}\sum_{i=0}^{\infty} (i+1)
\int dk_z \, \Theta( \bar{\mu} - \epsilon_{i k_z} ) \; ,  
\eeq
where $\epsilon_{i k_z} = 
\hbar \omega_{\bot} (i + 1) + \hbar^2 k_z^2/(2m)$. 
The term with $i=0$ gives the density of a strictly 
1D Fermi gas (only the lowest single-particle mode 
of the harmonic oscillator is occupied), while the terms with 
$i>0$ take into account occupation of the excited single-particle 
modes of the harmonic oscillator. After integration, 
the density $n_1$ can be written as 
\beq 
n_1 = \sum_{i=0}^{I[{\mu\over \hbar\omega_{\bot}}]}  
\, (i+1) \, {2^{3/2}\over \pi a_{\bot} } 
\sqrt{ {\mu \over \hbar \omega_{\bot}} - i } \; , 
\label{n1mu}
\eeq
where 
$a_{\bot}=\sqrt{\hbar /(m\omega_{\bot})}$ is the 
characteristic length of the transverse harmonic potential, 
$\mu =\bar{\mu} - \hbar \omega_{\bot}$ is the chemical potential
measured with respect to the ground state energy and  
$I[x]$ is the integer part of $x$, so that the summation 
extends from the ground state to the last higher occupied state, 
which may be  only partially occupied. 

In the upper panel of Fig. 1 we plot the chemical potential $\mu$ versus 
the transverse density $n_1$ obtained by using Eq. (\ref{n1mu}). 
The Fermi gas is strictly 1D only for
$0\le n_{1} < 2^{3/2}/(\pi a_{\bot})$, i.e. for 
$0 \le \mu < \hbar \omega_{\bot}$. In this case from 
Eq. (\ref{n1mu})  one finds
\beq
\mu = {\pi^2\over 8} \, \hbar\omega_{\bot} \,
(a_{\bot} n_1)^2 \; ,
\eeq
For $n_1 > 2^{3/2}/(\pi a_{\bot})$, i.e. for
$\mu > \hbar \omega_{\bot}$, several single-particle
states of the transverse harmonic oscillator are occupied
and the gas exhibits the 1D-3D crossover, becoming fully 3D when
$n_1 \gg 2^{3/2}/(\pi a_{\bot})$, 
i.e. when  
\beq
\mu = \left({15\pi\over 2^{7/2}}\right)^{2/5} \,
\hbar\omega_{\bot} \, (a_{\bot} n_1)^{2/5} \gg\hbar\omega_{\bot}\; 
\label{figata-1D}
\eeq
as obtained from (\ref{n1mu}).
\begin{figure}[tbp]
\begin{center} 
{\includegraphics[height=3.3in,clip]{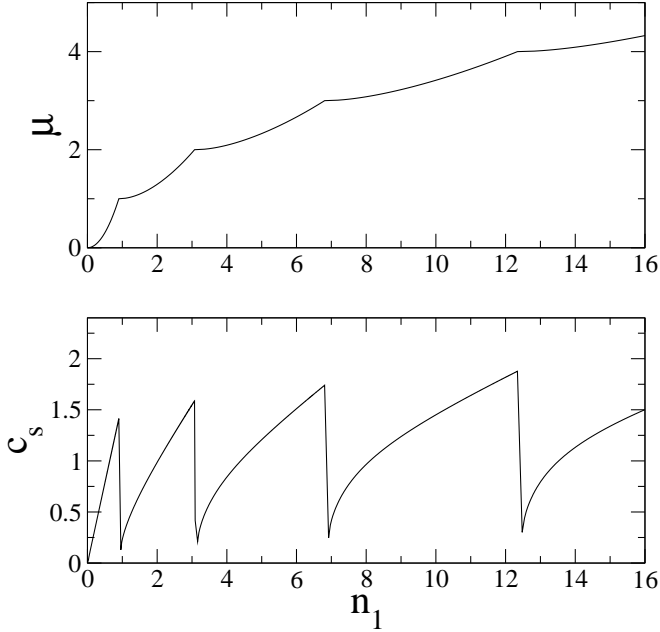}}
\end{center}
\caption{Chemical potential $\mu$ 
and first-sound velocity $c_s$ 
as a function of the axial density $n_1$ for an axially uniform Fermi 
gas with a radial harmonic confinement of frequency $\omega_{\bot}$. 
Chemical potential in units of $\hbar \omega_{\bot}$, 
sound velocity in units of 
$a_{\bot}\omega_{\bot}$, and density in units of 
$1/a_{\bot}$, where $a_{\bot}=\sqrt{\hbar/(m\omega_{\bot})}$. } 
\label{Fig1} 
\end{figure}

Since the Fermi temperature $T_F$ is related to the chemical potential
by the simple equation $k_BT_F=\mu$, where $k_B$ is Boltzmann's 
constant, our results are valid if 
the temperature $T$ of the system satisfies
the condition $T \ll \mu/k_B$. For instance,
in the case of $^{40}K$ atoms, setting 
$a_{\bot}=0.25$ $\mu$m, we find the harmonic transverse 
frequency $\omega_{\bot}\simeq 25$ kHz, which is well 
below the maximal confining transverse frequency obtained with 
permanent-magnetic atoms chips fo the study of long and thin 
atom clouds \cite{garcia}.  
The Fermi temperature of a strictly 1D gas of $^{40}$K atoms 
is then $T_F\simeq 130$ nKelvin. 

To analyze a sound wave that travels in the axial direction
it is important to determine 
the collision time $\tau_c$ of the gas. According to Bruun {\it et al.}
\cite{bruun} and Gupta {\it et al.}\cite{gupta}, 
if the local Fermi surface is not strongly deformed then 
$\tau_c = \tau_0 (T/T_F)^2$, where $\tau_0=1/(n \sigma v_F)$
with $\sigma$ the scattering cross-section and $n$ the
3D density. Instead, if the local Fermi surface is strongly deformed
as in our case, then the collision time is simply given
by $\tau_c=\tau_0$.\cite{gupta} 
It is easy to find that 
in our quasi-1D geometry $\tau_0 \omega_{\bot} 
\simeq 1/(n_1 a_s)^2$, 
where $a_F$ is the s-wave scattering length between fermions 
with opposite spins. 
As said above, the wave propagates in the collisionless regime 
if $\omega \tau_c \gg 1$ and in the collisional regime
if $\omega \tau_c \ll 1$.
The collision time $^{40}K$ atoms 
under a transverse harmonic 
confinement of frequency $\omega_{\bot} \sim 25$ kHz, 
is $\tau_c \sim 10^{-2}$ sec and therefore the hydrodynamic 
regimes will be obeyed by waves with frequencies 
up to $\omega \simeq 10^2$ Hz. 

In the collisional regime, 
by using the thermodynamic formula which relates  
the compressibility to the chemical potential 
\cite{landau,pines,fetter,negele} we immediately find that the axial first 
sound velocity $c_s$ of the Fermi gas is 
\beq
c_s = \sqrt{ {n_1\over m} 
{\partial \mu\over \partial n_1} } \; . 
\label{sound-c1}
\eeq 
This formula, supplemented by Eq. (\ref{n1mu}), enables us 
to determine the behavior of $c_s$ versus $n_1$, as shown in the lower 
panel of Fig. 1. The figure clearly displays shell effects, namely 
jumps of the first sound velocity $c_s$ 
when the atomic fermions occupy a new axial harmonic mode.
These shell effects could be tested experimentally. 
In fact, due to the  large anisotropy, transverse modes are decoupled 
from the axial sound modes.\cite{maruyama} 

As previously stressed the Fermi gas is strictly 1D only for 
$0\le n_{1} < 2^{3/2}/(\pi a_{\bot})$, i.e. for 
$0 \le \mu < \hbar \omega_{\bot}$. In this case from 
Eqs. (\ref{n1mu}) and (\ref{sound-c1}) one finds 
\beq 
c_s = {\pi\over 2} 
\, a_{\bot} \omega_{\bot} \, a_{\bot} n_1 \; . 
\label{scopa} 
\eeq 
Note that the asymptotic formula (\ref{scopa}) 
is exactly that discussed by Minguzzi 
{\it et al.}\cite{minguzzi} 
Instead, for $n_1 \gg 2^{3/2}/(\pi a_{\bot})$, i.e. 
for $\mu \gg \hbar \omega_{\bot}$, 
the Fermi gas becomes 3D and Eqs. (\ref{n1mu}) and (\ref{sound-c1}) 
one gets 
\beq 
c_s = \sqrt{2\over 5} 
\left({15\pi\over 2^{7/2}} \right)^{1/5} \, a_{\bot}\omega_{\bot} \, 
(a_{\bot} n_1)^{1/5} \; . 
\label{scopa1} 
\eeq 
The asymptotic result (\ref{scopa1}) coincides with that found 
by Capuzzi {\it et al.}\cite{capuzzi1} 
We stress that in strictly one-dimensional Fermi system the hydrodynamical 
approach gives the full spectrum of 
collective density fluctuations and the zero-sound velocity 
coincides with the first sound velocity\cite{wen}. 

To determine the conditions under which 
the first sound can be detected with 
a quasi-1D two-component Fermi gas, we consider again 
$^{40}K$ atoms with scattering length 
$a_F \simeq 150\cdot 10^{-10}$ m. 
In a cylindrical configuration 
with $L=1$ mm and $a_{\bot}=0.25$ $\mu$m 
we get the collision time 
$\tau_c \simeq 1.1\cdot 10^{-2}$ sec, and the sound velocity 
$c_s\simeq a_{\bot} \omega_{\bot} = 6$ mm/sec. 
The Fermi system is strictly 1D if the axial density $n_1$ 
does not exceed the value $n_1\simeq 4$ atoms/$\mu$m$^{-1}$. 
The characteristic wave length of collision is 
$\lambda_0 = c_s \tau_c \simeq 69$ $\mu$m. 
The condition for collisional regime is that the wave length 
$\lambda = c_s 2\pi/\omega$ of the sound wave is 
larger than $\lambda_0$. 
Thus, pertubing the axially uniform Fermi 
gas with a blu-detuned laser beam with a width of, 
for instance, $\simeq 150$ $\mu$m 
one produces two counter-propagating axial waves moving 
at the first-sound velocity $c_s$.  

\section{Conclusions}

We have shown that the first-sound velocity gives a clear 
signature of the dimensional crossover of a two-component 
normal Fermi gas. Our calculations suggest 
that the dimensional crossover induces shell effects, which 
can be detected as jumps in the first-sound velocity. 
We have discussed the experimental achievability 
of first sound by using gases of $^{40}$K atoms, 
finding that the collisional regime requires severe geometric 
and thermodynamical constraints. 
Finally, it is relevant to stress that in a Fermi system 
the sound propagates also in the collisionless regime due 
to mean-field effects. 
In this regime the velocity of (zero) sound can be determined by using 
the Boltzmann-Landau-Vlasov kinetic equation of the 
phase-space Wigner distribution function. This important issue 
will be considered elsewhere. 

This work has been partially supported by Fondazione CARIPARO. 
L.S. has been partially supported by GNFM-INdAM. 


\end{document}